\documentclass[aps,twocolumn,prl]{revtex4}
\usepackage{amsfonts}
\usepackage[final,hiresbb]{graphicx}
\usepackage{amsmath}
\usepackage{amssymb}
\usepackage{amstext}

\begin{document}
\preprint{SUBMITTED TO PHYSICAL REVIEW B}
\title{Toroidal Carbon Nanotubes with Encapsulated Atomic Metal Loops}
\author{Mark T. Lusk and Nathaniel Hamm}
\email{mlusk@mines.edu}
\affiliation{Department of Physics, Colorado School of Mines}
\begin{abstract}
Toroidal carbon nanotubes can serve as hosts for encapsulated loops of atomic
metal wires. Such composite structures have been analyzed using density
functional theory for a semiconducting C$_{120}$ torus encapsulating chains of
Fe, Au and Cu atoms. The sheathed metal necklaces form a zigzag structure and
drops the HOMO/LUMO bandgap to less than 0.1 eV. The iron composite is
ferromagnetic with a magnetic moment essentially the same as that of bcc iron.
The azimuthal symmetry of these toroidal composites suggests that they may
offer novel elecromagnetic properties not associated with straight,
metal-encapsulated carbon nanotubes.

\end{abstract}
\date{July 31, 2007}
\maketitle

\section{Introduction}

\begin{figure}
[ptb]
\begin{center}
\includegraphics[
height=2.8176in,
width=3.039in
]%
{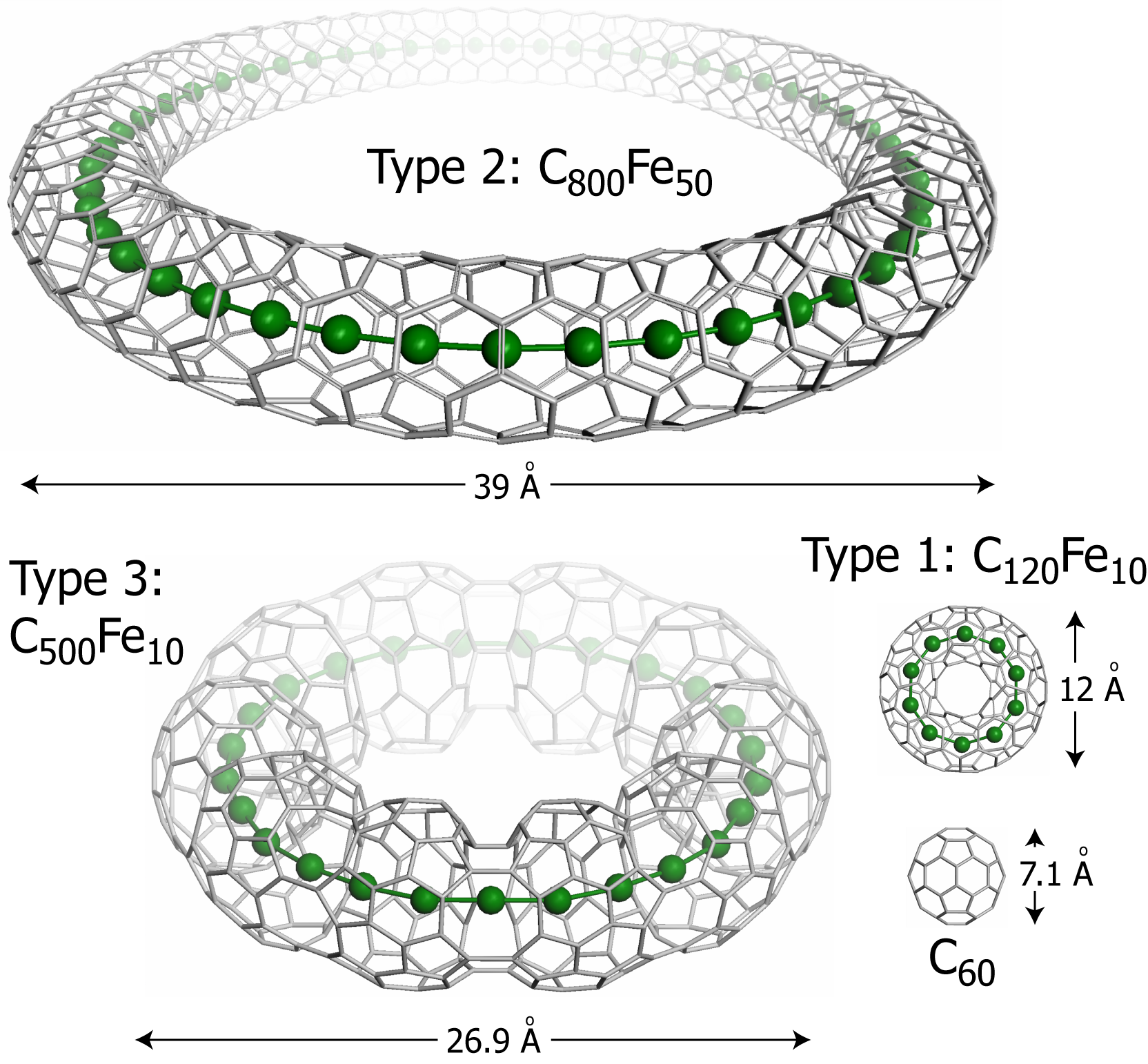}%
\caption{Atomic chains of Fe atoms encapsulated within three types of toroidal
carbon nanotubes. A bucky ball is shown at lower right for size comparison.
All structures are drawn to scale.}%
\label{figure_1}%
\end{center}
\end{figure}

Fullerene-based toroids combine the novel characteristics of carbon nanotubes
with electromagnetic and mechanical properties unique to closed-ring
geometries. An intriguing possibility, examined in this paper and shown in
Figure \ref{figure_1}, is that atomic chains of metal can be encapsulated
within such toroidal carbon nanotubes (TCNs). The carbon torus would serve as
a stabilizing sheath for the atomic metal necklaces resulting in rigid, stable
structures at finite temperatures with novel paramagnetic or ferromagnetic
properties properties. Beyond applications that exploit the properties of
independent rings, assemblies of such ferromagnetic tori might be expected to
self-assemble into columnar structures with unusual macroscopic behavior. The
electronic structure and mechanical stability of these robust atomic necklaces
are investigated using density functional theory (DFT) with the objective of
establishing ground state geometries, binding energies and magnetic moments.%

At least four types of single-wall fullerene toroids have been considered
since the idea was first proposed fifteen years ago\cite{PhysRevB.46.1933}.
Two variations of \textit{Type 1} TCNs exploit the curvature induced by
pentagonal and septagonal defects to generate polygonal graphitic
tori\cite{PhysRevB.46.1933},\cite{PhysRevB.47.12908} and rings as small as
$C_{120}$ have been considered\cite{PhysRevB.47.1703}. The symmetric
\textit{Type 2} TCNs are toroidal polyhexes\cite{kurby.Faraday.1993} created
by joining the ends of single-wall carbon nanotubes\cite{PhysRevB.57.14886}.
Perhaps the most technologically promising category of toroid is derived from
$C_{60}$ molecules that are joined with pentagonal and septagonal rings and
looped into undulated rings\cite{manzo:NanoLtrsi2004}%
,\cite{terrones:SolStateSci2006}. These are referred to here as\textit{\ Type
3} toroids. \ Finally, Haeckelite tubules\cite{PhysRevLett.84.1716}, based on
a two-dimensional bravais lattice with $5$, $6$ and $7$-member rings, can also
be joined at the ends to make the \textit{Type 4}
TCNs\cite{manzo:NanoLtrsi2004},\cite{terrones:SolStateSci2006}. The first two
types of TCNs are theoretical constructs that are thought to be mechanically
stable. It has been estimated that TCNs of \textit{Type} 1 are in fact more
stable than C$_{60}$ molecules\cite{PhysRevB.47.1703}, while \textit{Type} 2
TCNs are expected to be stable provided they have a sufficiently large
toroidal diameter\cite{PhysRevB.57.14886},\cite{PhysRevB.72.085416}. Carbon
toroids of \textit{Types} 3 and 4, on the other hand, have been realized
experimentally and are demonstrably stable\cite{manzo:NanoLtrsi2004}%
,\cite{terrones:SolStateSci2006}. \ 

For all toroids, delocalized $\pi$ electrons can be induced into ballistic,
persistent motion by the application of a magnetic field perpendicular to the
plane of the torus. The resulting orbital magnetic response of tori can be
substantial and has been investigated using tight binding theory with the
magnetic vector potential accounted for using London
theory\cite{LeRadium.8.397},\cite{PhysRevLett.74.1123}. \textit{Type} 1
toroids are predicted to exhibit a large diamagnetic response
\cite{haddon:Nature1991},\cite{haddon:Nature1997},\cite{PhysRevB.57.14886},
while a class of conducting \textit{Type} 2 TCNs are expected to be able to
generate colossal paramagnetic moments \cite{PhysRevB.57.6731}%
,\cite{PhysRevB.57.14886},\cite{PhysRevLett.88.217206}. More recently, carbon
toroids of \textit{Types} 3 and 4 TCNs have been shown experimentally to have
paramagnetic behavior consistent with tight binding model predictions
\cite{manzo:NanoLtrsi2004}. The septagonal and pentagonal rings in these
structures are predicted to result in ferromagnetic behavior at sufficiently
low temperatures in the absence of any flaws. All types of TCNs are expected
to exhibit Aharanov-Bohm effects\cite{haddon:Nature1997}.

Paralleling activity associated with toroidal fullerene structures have been
advances in the fabrication and analysis of isolated atomic chains of metal
atoms. Metal contacts created with a scanning tunneling microscope can be
slowly pulled apart to extrude an atomic chain of Au
atoms\cite{yanson:Nature1998},\cite{ohnishi:Nature1998}. Stable monatomic
chains of Au atoms can support currents of $80$ $\mu A$--clear evidence of
ballistic transport in this alternate manifestation of one-dimensional
conductors\cite{yanson:Nature1998}--and the quantum conductance in these
metallic atomic chains has also been measured\cite{ohnishi:Nature1998}. As
subsequent DFT analysis of atomic chains of Au atoms, intended to address
observed anomalies, indicated that zigzag structures are more stable than a
linear chains as a result of reduced transverse kinetic energy of
electrons\cite{PhysRevLett.83.3884}. Such configurations are most likely
metastable since previous investigations have identified polyhedral ground
state structures for small clusters of Fe and Au
atoms\cite{rollmann:CompMatSci2006},\cite{Jain.16.2005}.

Iron structures sheathed within straight single wall carbon nanotubes have
been studied both computationally and
experimentally\cite{rahman:JPhysCondMat2004},\cite{PhysRevB.73.125435}. A DFT
analysis of atomic chains of Fe within single-wall carbon nanotubes concluded
that the spacing between Fe atoms and their nearest C neighbors determined the
character of the Fe wire\cite{rahman:JPhysCondMat2004}. The Fe-C bond length
of $1.9$ $\mathring{A}$ associated with $(3,3)$ tubes resulted in a composite
with semi-conductor character and no magnetic moment. On the other hand, an
Fe-C bond length of $2.1$ $\mathring{A}$ associated with $(4,4)$ tubes
resulted in a composite with the metallic character and magnetic moment of
$2.6\mu_{B}$ per atom\cite{rahman:JPhysCondMat2004}. The geometry of such
tubes precludes the generation of axial persistent currents and a colossal
paramagnetic response, but they might be used to help predict the conduction
and ferromagnetic nature wires encapsulated within TCNs.

\begin{figure}
[ptb]
\begin{center}
\includegraphics[
height=4.3552in,
width=2.2866in
]%
{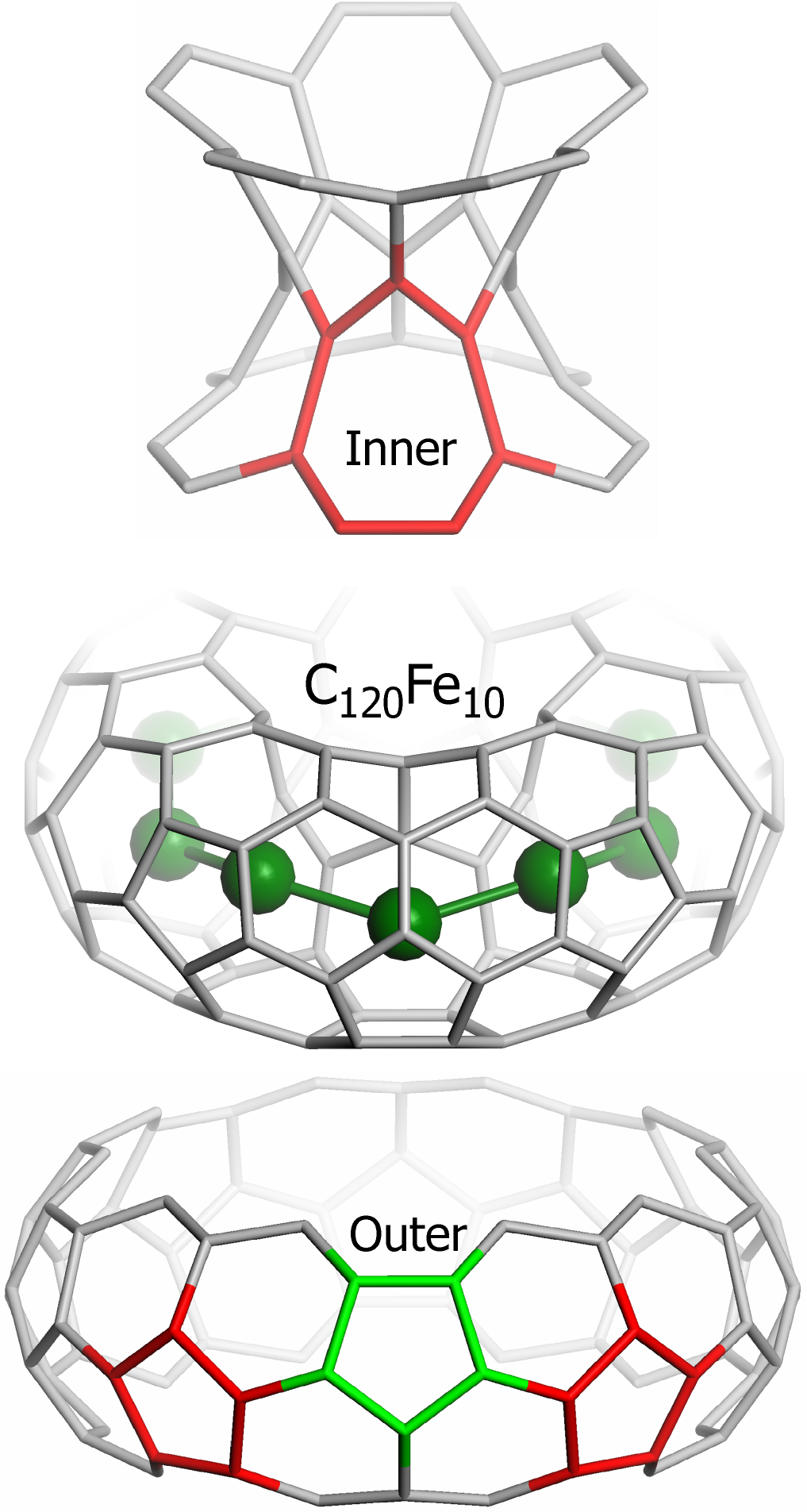}%
\caption{Break-out view of the inner and outer structures of the C$_{120}%
$Fe$_{10}$ molecule with shading used to highlight polygonal defect types.}%
\label{c120_inner_outer}%
\end{center}
\end{figure}

The present work brings together the unusual behavior of toroidal carbon
nanotubes and atomic metallic chains within a simple setting: a \textit{Type
1} C$_{120}$ torus with D$_{5d}$ symmetry\cite{PhysRevB.47.12908}
encapsulating closed, monatomic chains of Au$,$ Fe, and Cu atoms. The iron
necklace is of particular importance because of the possibility that it has a
ferromagnetic nature at finite temperatures. The gold necklace was also
considered, though, because of the focus that Au has received in atomic metal
wires\cite{yanson:Nature1998},\cite{ohnishi:Nature1998}. Analysis was Cu atoms
was thought to offer an intermediate composite since it is a noble metal, like
Au, and a Period 4 metal like Fe. The \textit{Type} 1 TCN chosen for analysis
is a polygonal C$_{120}$ torus with $D_{5d}$ symmetry--chosen for its
predicted stability and minimal size\cite{PhysRevB.47.12908}. The C$_{120}$
torus is composed of an inner surface of ten septagons and an outer surface of
ten pentagons patched with twenty hexagons as shown in Figure
\ref{c120_inner_outer}. Encapsulated atomic loops composed of ten metal atoms
were considered in order to exploit maintain the computational expedient of
D$_{5d}$ symmetry while creating M-M bond lengths on the order those
calculated for isolated metal chains. Density functional theory (DFT) was used
to determine the ground state configurations and to quantify M-M bond length,
minimum C-M bond length, magnetic moment, density of states, conduction
character, binding energy of the encapsulated metal atoms.%

\section{Method}

All calculations were performed with the real-space, numerical atomic orbital,
DFT code, \text{DMOL}$^{3}$\cite{JChemPhys.113.7756}. A norm conserving, spin
unrestricted, semi-core pseudopotential approach\ was employed with electron
correlation accounted for using the Perdew--Wang generalized gradient
approximation (GGA)\cite{PhysRevB.45.13244}. Double numerical basis sets were
used with $4p$ polarization functions for Fe atoms and $3d$ polarization
functions for C atoms. A smearing value of $0.272$ eV was used for all structures.

As a check on the method, an Fe dimer was determined to have a ($^{7}%
\Delta_{u}$) collinear ferromagnetic ground state with a bond length of $2.00$
$\mathring{A}$ and magnetic moment of $7.0\mu_{B}$ by Hirshfeld
analysis--consistent with both experimental and other DFT
results\cite{reddy:068301}. As a further check, the atomic spacing, binding
energy and magnetic moment of bcc Fe was evaluated to be $2.48$ $\mathring{A}%
$, $-4.89\ $ eV and and $2.22\mu_{B}$--reasonable matches to the experimental
values of $2.47$ $\mathring{A}$, $-4.32\ $ eV and $2.22\mu_{B}$ per
atom\cite{ackland:PhilMag1997},\cite{Kittel},\cite{PhysRevB.49.6012}. DFT
values of $2.76\ $ eV and $2.08\mu_{B}$ per atom were obtained
elsewhere\cite{garcia:EuroJPhys2004}.

\section{Results}

\subsection{Free Standing Metal Chains}

\begin{figure}
[ptb]
\begin{center}
\includegraphics[
height=1.1831in,
width=2.2866in
]%
{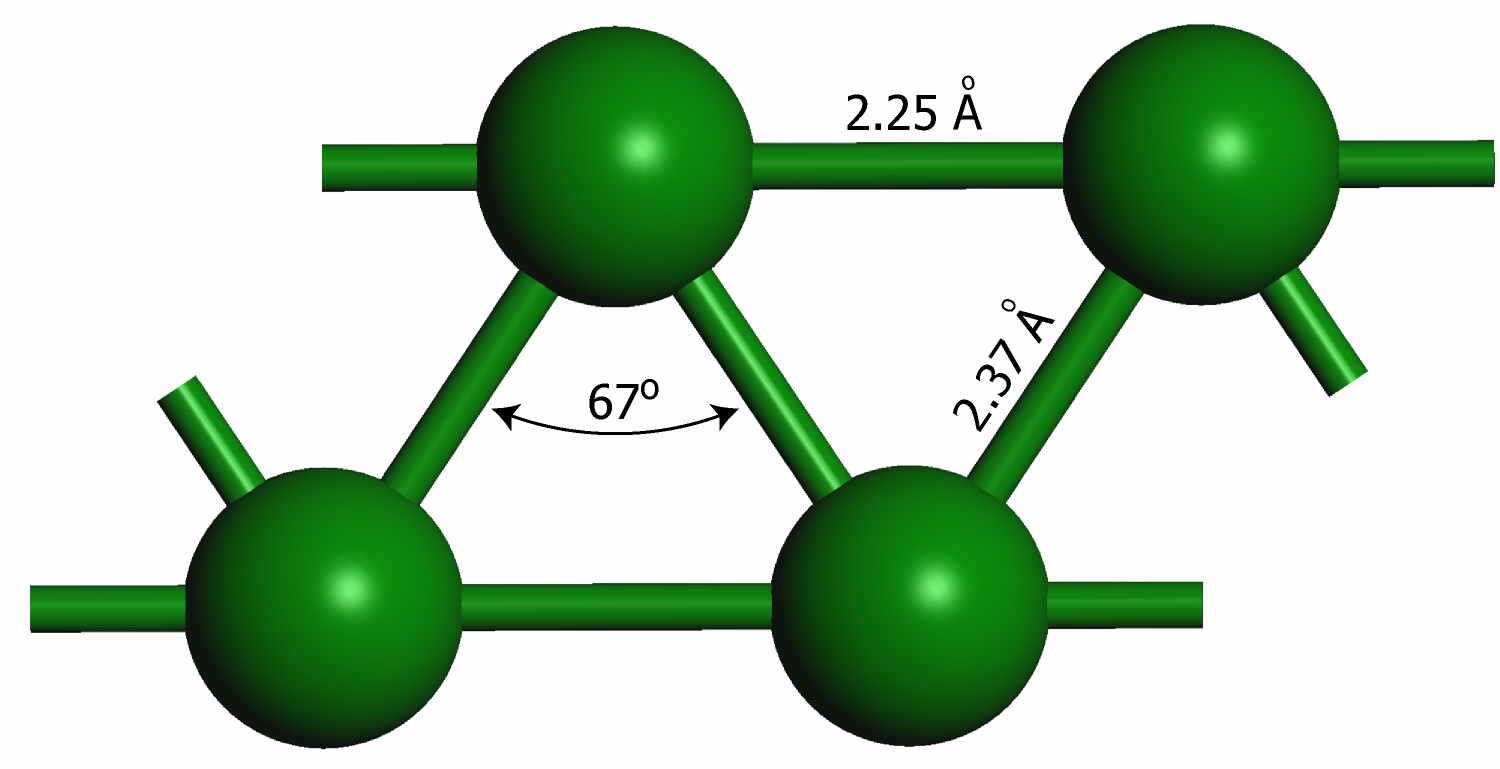}%
\caption{Ground state structure for a periodic chain of Fe atoms. }%
\label{figure_2}%
\end{center}
\end{figure}

Periodic structures of pairs of free standing Fe, Au and Cu atoms were
analyzed in order to understand the character of metallic chains prior to
encapsulating them in carbon toroids. The ground state configuration for the
Au chain was found to be a zigzag structure with a bond length of $2.83$
$\mathring{A}$ and an angle between adjacent atoms of $57{{}^\circ}$. There is also a local equilibrium of higher energy at $134{{}^\circ}$, and this is close to that obtained by S\'{a}nchez-Portal et al. who
identified an equilibrium angle of $131{{}^\circ}$\cite{PhysRevLett.83.3884}. Analogous results, summarized in Tables 2 and 3,
were obtained for Fe and Cu chains, and all three chains exhibit ground state
bending angles of approximately $60{{}^\circ}$; an affinity for symmetric, three-atom bonding is optimum within the
geometric constraints imposed. As shown in Figure \ref{figure_2}, such
structures amount to two-dimensional, triangular lattices structures rather
than linear chains. Key property data is summarized in the tables. All
positive eigenvalues of the Hessian matrices associated with periodic cells of
four atoms indicate at least linearized stability at $0$ $K$ for all three
chains. A Jahn-Teller distortion, manifested in both bond angle and length,
was noted for metallic strands of eight atoms or more, and this is consistent
with previous ab initio investigations\cite{Yanson.1975}.%

\subsection{Atomic Metal Wires within SWCNT}

\begin{figure}
[ptb]
\begin{center}
\includegraphics[
height=2.3307in,
width=2.2866in
]%
{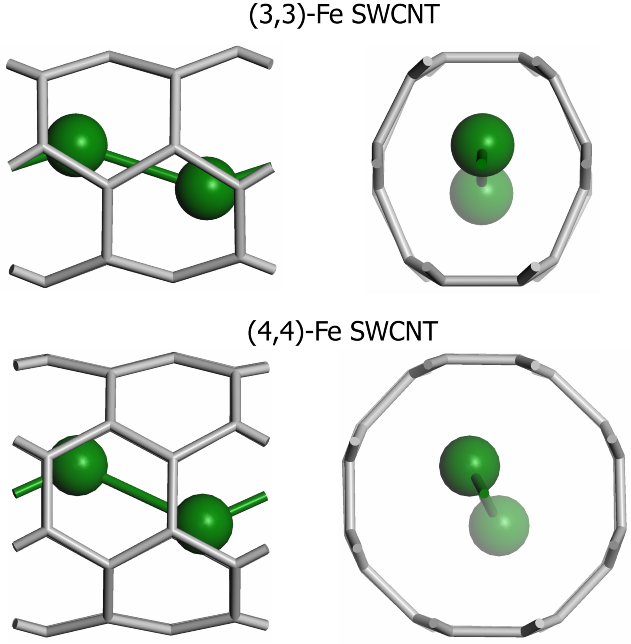}%
\caption{Ground state structures of $(3,3)$ and $(4,4)$ SWCNT embedded with
atomic Fe wire.}%
\label{figure_3}%
\end{center}
\end{figure}

\begin{figure}
[ptb]
\begin{center}
\includegraphics[
height=2.3428in,
width=2.2866in
]%
{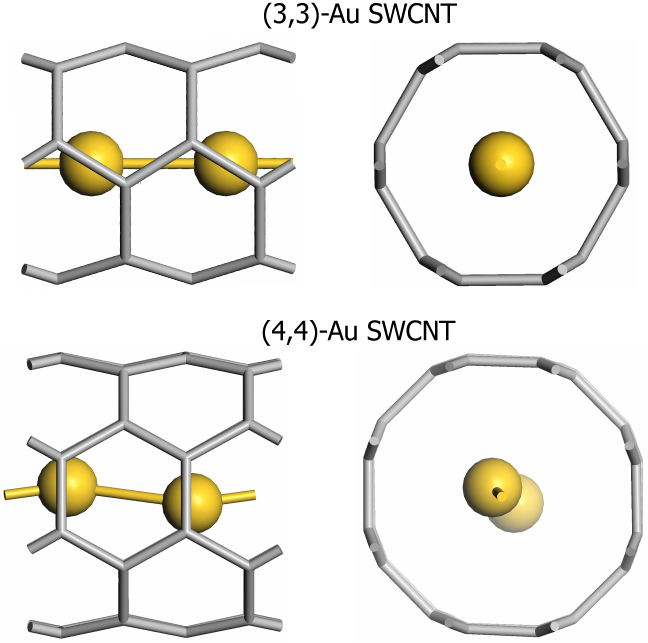}%
\caption{Ground state structures of $(3,3)$ and $(4,4)$ SWCNT embedded with
atomic Au wire.}%
\label{cnt_au_composite}%
\end{center}
\end{figure}

Single-strand atomic wires of metal atoms encapsulated within single-wall
carbon nanotubes (SWCNT) are of interest on their own and can be used to
anticipate the electronic structure and stability of wires encapsulated within
carbon tori. The tube diameter of the C$_{120}$ torus is between that of
$(3,3)$ and $(4,4)$ SWCNT, and so these fullerenes were chosen for more
detailed analysis with metal wires.\ Previous DFT studies considered atomic
wires of Fe within SWCNT\cite{rahman:JPhysCondMat2004},\cite{Kisaku.2005}. The
analysis for $(n,n)$ tubes considered only one metal atom per unit cell,
though, and therefore did not allow for the possibility of the metal chain
forming a zigzag configuration. Since the toroidal structure should allow for
such corrugation, the SWCNT analysis was performed for double-cell, periodic
structures with both Fe, Au and Cu wires. The resulting equilibrium
configurations for are shown in Figures \ref{figure_3}, \ref{cnt_au_composite}
and \ref{cnt_au_composite} with property data summarized in Tables 1 through
3. As expected, the encapsulated wires do exhibit a zigzag structure although
the carbon sheath flattens the wires relative to their free standing
configuration. For the $(4,4)$ tube, the Fe wire causes a diametral tube
strain of $0.94\%$ while the Au wire causes a larger diametral expansion with
$2.44\%$. Reassuringly, the Cu wire causes a dilation between these
two--$2.1\%$. In the case of the smaller $(3,3)$ tube, the diametral expansion
of the Au tube is $8.7\%$, while the Fe tube shows a non-uniform distortion
with bi-lateral strains of $8.2\%$ and $-5.2\%$. The Cu tube is uniformly
dilated, like Au case, with a strain of $3.4\%$. The allowance for zigzagging
within the tubes significantly impacts the electronic structure of the
composite. If a periodic cell half as long is considered, so that there is
only one Fe atom per period, the bandgap is $0.87\ $ eV instead of $0.48\ $
eV. The first number is consistent with the $0.88\ $ eV obtained
previously\cite{Kisaku.2005}. The magnetic moment of both straight and
zigzagging chains of Fe atoms were extinguished within $(4,4)$ CNTs. An
earlier analysis on straight, sheathed chains reported a magnetic moment of
$2.6\mu_{B}$ per Fe atom\cite{Kisaku.2005}.%

%

%

\begin{figure}
[ptb]
\begin{center}
\includegraphics[
height=2.4206in,
width=2.2866in
]%
{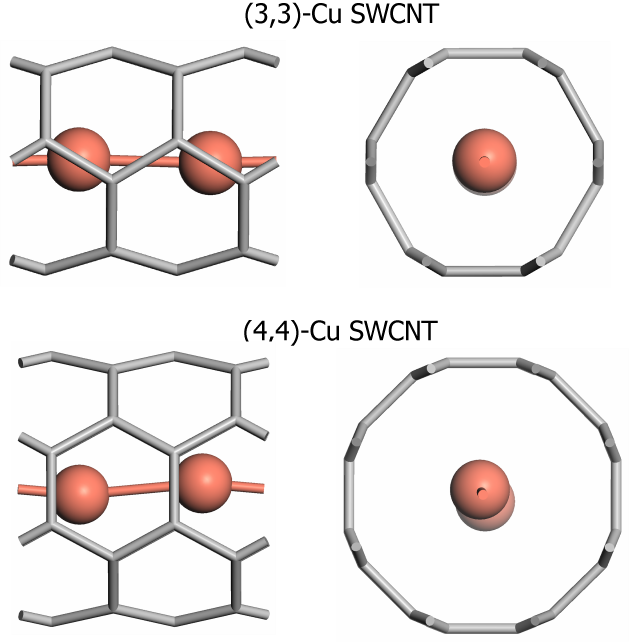}%
\caption{Ground state structures of $(3,3)$ and $(4,4)$ SWCNT embedded with
atomic Cu wire.}%
\label{cnt_cu_composite.png}%
\end{center}
\end{figure}

\subsection{C$_{120}$ Torus}

The C$_{120}$ torus was first analyzed in the absence of atomic wire, and the
energy per C atom was found to be less than that of C$_{60}$ by $0.09\ $ eV
(smearing) and $0.16\ $ eV (no smearing). This is in contrast to an earlier
tight binding calculation which predicted that the toroidal state would be of
lower energy\cite{PhysRevB.47.12908}. The difference is small, though, and the
C$_{120}$ torus should be very stable. A reactive molecular dynamics
code\cite{PhysRevB.72.085416} was used to estimate that the structure is
stable to at least $2000$ K. The torus has no intrinsic magnetic moment and
exhibits a bandgap of $0.91\ $ eV between its highest occupied and lowest
unoccupied molecular orbitals (HOMO/LUMO). For comparison, the HOMO/LUMO
bandgap of C$_{60}$ was found to be $1.5\ $ eV--close to the experimental
measurement of $1.6\ $ eV\cite{C60.1991}.

\subsection{Atomic Metal Necklaces within a C$_{120}$ Torus}

\begin{figure}
[ptb]
\begin{center}
\includegraphics[
height=3.1704in,
width=2.2866in
]%
{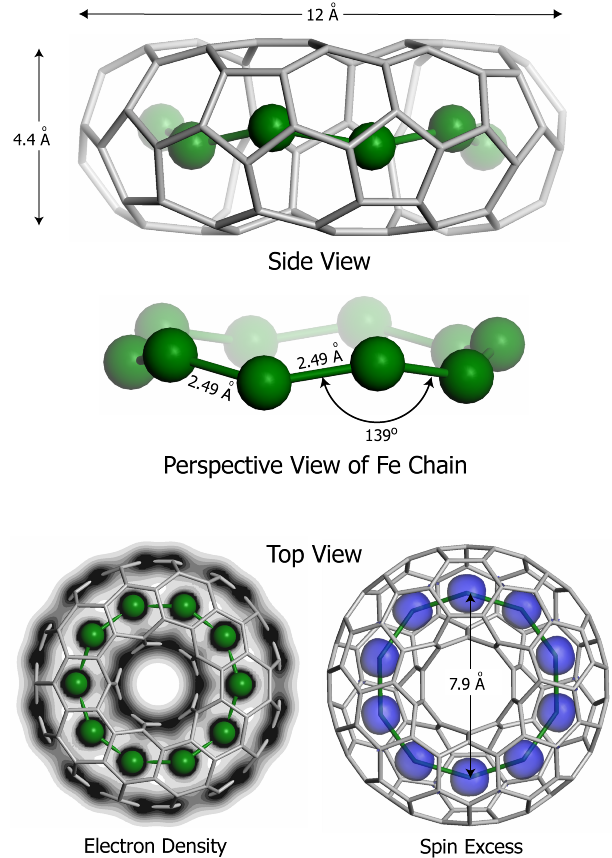}%
\caption{Ground state structure of C$_{120}$Fe$_{10}$. The axial positions of
the Fe atoms alternate between $\pm0.21$ $\mathring{A}$. Also shown are the
electron density cross-section (center) and spin excess isosurface (far
right).}%
\label{figure_4}%
\end{center}
\end{figure}

\begin{figure}
[ptb]
\begin{center}
\includegraphics[
height=1.2756in,
width=2.2866in
]%
{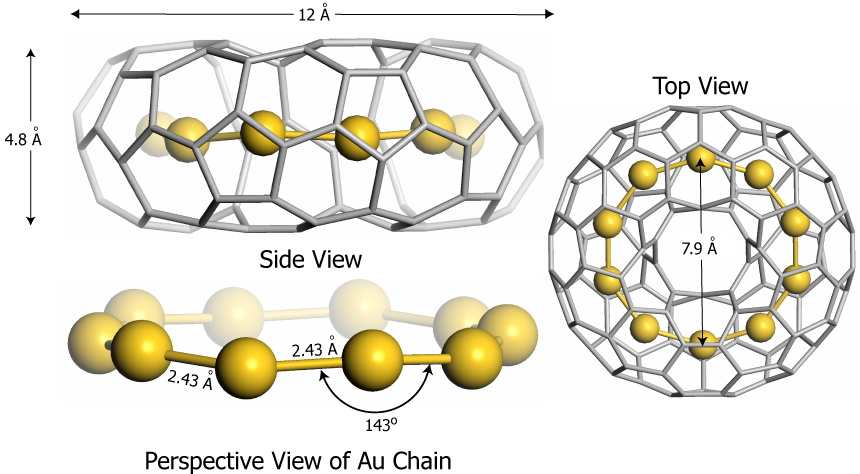}%
\caption{Ground state structure of C$_{120}$Au$_{10}$. The axial positions of
the Au atoms alternate between $\pm0.076\mathring{A}$. Also shown are the
electron density cross-section (center) and spin excess isosurface (far
right).}%
\label{c120_au10_equil}%
\end{center}
\end{figure}

\begin{figure}
[ptb]
\begin{center}
\includegraphics[
height=1.3526in,
width=2.2857in
]%
{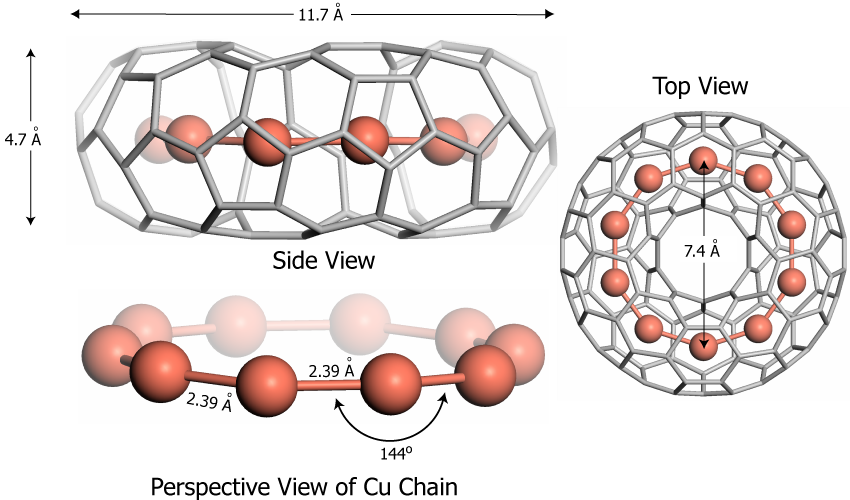}%
\caption{Ground state structure of C$_{120}$Cu$_{10}$. The axial positions of
the Cu atoms alternate between $\pm0.025\mathring{A}$. Also shown are the
electron density cross-section (center) and spin excess isosurface (far
right).}%
\label{c120_cu10_equil}%
\end{center}
\end{figure}

The ground state was obtained for M$_{10}$ loops encapsulated within the
C$_{120}$ torus with the resulting structures shown in Figures \ref{figure_4},
\ref{c120_au10_equil} and \ref{c120_cu10_equil}. The associated properties are
summarized in Tables 1-3. Consistent with the ground state morphology of free
standing metallic chains, the encapsulated necklace of Fe atoms forms a zigzag
structure which is highlighted in the bottom of the figure. The minimum Fe-C
bond length was found to be between that of the $(3,3)$ and $(4,4)$ SWNTs. The
toroidal composite has a HOMO/LUMO bandgap of less than 0.1 eV and has a
magnetic moment of $2.28\mu_{B}$ per atom--approximately the same as that
associated with bcc Fe. The spatial distribution of this magnetic moment is
shown in Figure \ref{figure_4} through an isosurface plot of the spin excess.
At zero temperature, the Fe atoms have aligned spin axes and so the composite
is ferromagnetic. In contrast, neither of the straight nanotubes with Fe
showed any intrinsic moment because of stronger Fe-C bonding for the tubes.

Encapsulation of an Au$_{10}$ necklace was found to be energetically
unfavorable, while the Cu$_{10}$ encapsulation is an energetically neutral
process. As indicated in Tables 2 and 3, these results are consistent with
those associated with CNT encapsulations.%

%

%

%

\begin{table}[table1] \centering
\begin{tabular}
[c]{||l||c|c|c|c|c|c||}\hline\hline
{\small Structure} & {\small M-M} & {\small M-C} & {\small M-M} & {\small Gap}
& {\small Mom.} & {\small E}$_{\text{B}}$\\
& {\small \AA } & {\small \AA }{\tiny min} & {\small Angle} & {\small eV} &
$\mu_{B}${\small /M} & {\small eV/M}\\\hline
{\small Fe bcc} & {\small 2.49} & {\small -} & {\small -} & {\small 0.00} &
{\small 2.22} & {\small 4.89}\\\hline
{\small Fe chain} & {\small 2.35} & {\small -} &
\multicolumn{1}{|c|}{{\small 59}$%
{{}^\circ}%
$} & {\small 0.00} & {\small 3.25} & {\small 2.87}\\\hline
{\small (3,3)\thinspace-Fe} & {\small 2.59} & {\small 1.94} & {\small 144}$%
{{}^\circ}%
$ & {\small 0.48} & {\small 0.00} & {\small 1.75}\\\hline
{\small (4,4)-Fe} & {\small 2.77} & {\small 2.22} & {\small 126}$%
{{}^\circ}%
$ & {\small 0.00} & {\small 0.00} & {\small 1.95}\\\hline
{\small C}$_{\text{120}}${\small -Fe}$_{\text{10}}$ & {\small 2.49} &
{\small 2.04} & {\small 139}$%
{{}^\circ}%
$ & {\small
$<$%
0.1} & {\small 2.28} & {\small 1.82}\\\hline\hline
\end{tabular}
\caption{Summary of properties for Fe structures. The bandgap is HOMO-LUMO for the toroidal molecule. \label{height}}%
\end{table}%
%

\begin{table}[table2] \centering
\begin{tabular}
[c]{||l||c|c|c|c|c|c||}\hline\hline
{\small Structure} & {\small M-M} & {\small M-C} & {\small M-M} & {\small Gap}
& {\small Mom.} & {\small E}$_{\text{B}}$\\
& {\small \AA } & {\small \AA }{\tiny min} & {\small Angle} & {\small eV} &
$\mu_{B}${\small /M} & {\small eV/M}\\\hline
{\small Au fcc} & {\small 2.88} & {\small -} & {\small -} & {\small 0.00} &
{\small 0.00} & {\small 3.05}\\\hline
{\small Au chain} & {\small 2.83} & {\small -} & {\small 57}$%
{{}^\circ}%
$ & {\small 0.00} & {\small 0.00} & {\small 2.07}\\\hline
{\small (3,3)-Au} & {\small 2.46} & {\small 2.34} & {\small 180}$%
{{}^\circ}%
$ & {\small 0.00} & {\small 0.00} & {\small -6.80}\\\hline
{\small (4,4)-Au} & {\small 2.52} & {\small 2.56} & {\small 156}$%
{{}^\circ}%
$ & {\small 0.00} & {\small 0.00\ } & {\small -0.12}\\\hline
{\small C}$_{\text{120}}${\small -Au}$_{\text{10}}$ & {\small 2.43} &
{\small 2.00} & {\small 143}$%
{{}^\circ}%
$ & {\small
$<$%
0.1} & {\small 0.00\ } & {\small -5.54}\\\hline\hline
\end{tabular}
\caption{Summary of properties for Au structures. The bandgap is HOMO-LUMO for the toroidal molecule. \label{height}}%
\end{table}%
%

\begin{table}[table3] \centering
\begin{tabular}
[c]{||l||c|c|c|c|c|c||}\hline\hline
{\small Structure} & {\small M-M} & {\small M-C} & {\small M-M} & {\small Gap}
& {\small Mom.} & {\small E}$_{\text{B}}$\\
& {\small \AA } & {\small \AA }{\tiny min} & {\small Angle} & {\small eV} &
$\mu_{B}${\small /M} & {\small eV/M}\\\hline
{\small Cu fcc} & {\small 2.56} & {\small -} & --- & {\small 0.00} &
{\small 0.00} & {\small 3.62}\\\hline
{\small Cu chain} & {\small 2.42} & {\small -} & {\small 59}$%
{{}^\circ}%
$ & {\small 0.00} & {\small 0.00} & {\small 2.17}\\\hline
{\small (3,3)-Cu} & {\small 2.46} & {\small 2.24} & {\small 177}$%
{{}^\circ}%
$ & {\small 0.00} & {\small 0.00} & {\small -0.83}\\\hline
{\small (4,4)-Cu} & {\small 2.47} & {\small 2.69} & {\small 168}$%
{{}^\circ}%
$ & {\small 0.00} & {\small 0.00\ } & {\small 1.09}\\\hline
{\small C}$_{\text{120}}${\small -Cu}$_{\text{10}}$ & {\small 2.39} &
{\small 1.92} & {\small 144}$%
{{}^\circ}%
$ & {\small
$<$%
0.1} & {\small 0.00\ } & {\small 0.01}\\\hline\hline
\end{tabular}
\caption{Summary of properties for Cu structures. The bandgap is HOMO-LUMO for the toroidal molecule. \label{height}}%
\end{table}%

The DFT investigation indicates that the C$_{120}$ torus with all three types
of metal\ necklaces is structurally stable at zero temperature. The TCN has a
strong affinity for Fe atoms ($1.82\ $ eV/Fe) and a neutral affinity for Cu
atoms, but the placement of a Au necklace within the torus would require an
energy input of $5.54\ $ eV per Au atom. This precludes the study of Au wires
within this setting. The $(3,3)$ and $(4,4)$ CNTs can be used to estimate the
affinity of C$_{120}$ toroids for metal atoms; the binding energy was found to
be midway between the CNT values for each of the three metal chains. This
allowed the prediction, in advance, that C$_{120}$ would exhibit a neutral
affinity for the Cu necklace. The presence of the metal atoms causes the TCN
composite to have a HOMO/LUMO bandgap of less than 0.1 eV in all three cases.
When encapsulated with an Fe necklace, the torus exhibits ferromagnetic
properties close to that of bcc iron, and it will be useful to determine to
what degree, if any, this persists at finite temperatures. It is not clear
what sort of chain structure and composite properties would result from the
encapsulation of chains with symmetry at odds with that of the host torus.

\section{Summary}

The\ azimuthal symmetry of encapsulated atomic metallic necklaces within
toroidal carbon sheaths offers the prospect of synthesizing composites with
novel electromagnetic properties that are not associated with straight,
metal-encapsulated CNTs. As a first step in this direction, a particularly
simple encapsulation was analyzed using density functional theory to determine
its ground state geometry, structural stability, binding energy, electrical
conductivity and magnetic moment. Of the four known classes of TCNs, a Type 1
C$_{120}$ molecule was chosen because it results in a computationally
tractable composite, and only ten metal atoms were considered in order to take
full computational advantage of the D$_{5d}$ symmetry of the torus. The TCN
exhibits a positive, neutral and negative affinities for Fe, Cu and Au atoms,
respectively, and metal atom encapsulation resulted in linearly stable
structures in all three cases. In contrast to Fe chains within $(3,3)$ and
$(4,4)$ CNT's, Fe atoms within the C$_{120}$-Fe$_{10}$ molecule exhibit a net
magnetic moment comparable to bcc Fe.

All toroidal nanotubes should all be capable of serving as sheathing hosts for
atomic metal necklaces with novel electromagnetic properties. The possibility
of encapsulating metal chains within the C$_{60}$-coalesced tori of
\textit{Types} 3 and 4 is of particular interest because these are structures
that have already been synthesized\cite{manzo:NanoLtrsi2004}%
,\cite{terrones:SolStateSci2006}. For \textit{Type} 3 TCNs, a reasonable
strategy might be to encapsulate Fe atoms within the C$_{60}$ molecules prior
to coalescing them into a toroid. Sufficiently long CNTs with embedded metal
atoms could also be deformed to synthesize Type 2 TCN composites. Both
approaches would admit the construction of alloyed necklaces as well---the
one-dimensional, atomic analog of powder metallurgy.

\newif\ifabfull\abfulltrue

\end{document}